# Electronic structure and electron energy-loss spectra of $Sr_{0.35}CoO_2$

R. J. Xiao, H. X. Yang, L. F. Xu, H. R. Zhang, Y. G. Shi and J. Q. Li*

Beijing National Laboratory for Condensed Matter Physics, Institute of Physics, Chinese Academy of Sciences, Beijing, 100080, China

Abstract

The electronic structure of $Sr_{0.35}CoO_2$, structurally analogous to the layered $Na_xCoO_2$, has been evaluated using the local spin density approximation (LSDA). We find that evident c-dispersions appear in both the $e'_g$ and $a_{1g}$ Co-derived bands, demonstrating the existence of a notable interplanar interaction in $Sr_{0.35}CoO_2$. The LSDA+U calculation reveals that the electronic structure, in particular the band splitting between the spin-up and spin-down electrons, changes evidently along with the increase of the effective on-site Coulomb interaction U. Analysis of theoretical and experimental electron energy-loss spectra (EELS) for the oxygen K-edge and the Co $L_{23}$-edge suggests that the on-site interaction (U) in $Sr_{0.35}CoO_2$ is less than 3eV which is noticeably weaker than the estimated value (from 5eV to 8eV) in the $Na_xCoO_2$ materials.

# 1. Introduction

Since the discovery of the remarkable thermoelectric behaviors in $Na_xCo_2O_4$ [1], an intense research interest has been stimulated in alkali-metal intercalated layered cobaltites due to their potential technological applications [2-4]. The recent observations of superconductivity in $Na_{0.33}CoO_2·1.3H_2O$ [5], charge ordering in $Na_{0.5}CoO_2$ [6] and complex magnetic structures in $Na_xCoO_2$ (x=0.67, 0.75) [7] have further heightened the interest in this kind of materials. Intercalation of divalent ions ($Ca^{2+}$, $Sr^{2+}$) into cobalt oxides was expected to improve the thermoelectric properties. For instance, $Sr_{0.35}CoO_2$ shows a Seebeck coefficient comparable to that of $Na_xCoO_2$ [8]; partial substitution of Ca for Na in $Na_xCoO_2$ also gave an enhancement in both Seebeck coefficient and electric resistivity [2, 9]. Structural investigations revealed that the intercalations of either Na (Sr) atoms or $H_2O$ molecules could make the local structure very complex; the intercalated atoms can be random with high mobility or crystallized in certain ordered states [10]. Recently, we have successfully prepared a series of samples with nominal composition of $Ca_xCoO_2$ (0.25≤x≤0.5) and $Sr_xCoO_2$ (0.25≤x≤0.4) by the low-temperature ion-exchange technique. Structural analysis by means of transmission-electron microscopy and x-ray diffraction suggested that the well-defined Sr-ordered state only appears at x ≅ 1/3 in $Sr_xCoO_2$ (0.25≤x≤0.4) system [11], materials with other Sr concentrations contain evident structural imhomogeneity and phase separation. Moreover, it is noted that the structure of $Sr_{0.35}CoO_2$ is modulated by Sr ordering which yields a superstructure of $3^{1/2}a \times 3^{1/2}a$ [11]. This specific superstructure was expected to have large effects on the charge ordering and magnetic structures as theoretically demonstrated by first principle calculations [12]. In this paper, we report on the electronic band structure of $Sr_{0.35}CoO_2$ calculated by LSDA and LSDA+U with $0 \leq U \leq 8eV$. Certain theoretical EELS are analyzed in comparison with the experimental ones.

## 2. Computational details and experimental methods

### 2.1. Theoretical approach

The crystal structure of $Sr_{0.35}CoO_2$ used in our calculation was based on the x-ray diffraction result as reported by R. Ishikawa et al [8]. This phase has the lattice parameters of a=2.820Å and c=11.525Å with the space group of $P6_3/mmc$. The partially occupied Sr sites were simulated by virtual crystal approximation [13]. The full potential linearized augmented-plane-wave+local orbital (APW+LO) were used as implemented in WIEN2k code and its LDA+U extension [14, 15]. Self-consistency was carried out on a 16×16×3 mesh containing 60 k-points in the irreducible Brillouin zone (BZ). $R_{mt}K_{max}$ was set to 7.0 to determine the basis size. The parameter for exchange interaction (J) in the LSDA+U calculation were taken as J=0.965eV [16]. The core-loss EELS were calculated using TELNES program of the WIEN2k distribution [17].

### 2.2. Experiment

Polycrystalline materials of $Sr_{0.35}CoO_2$ were prepared by the low-temperature ion exchange reaction from the ɣ-$Na_{0.7}CoO_2$ precursor prepared by conventional solid-state reaction [8, 11]. The measurements of EELS were performed on a Tecnai F20 transmission electron microscope equipped with a post column Gatan imaging filter. The energy resolution in the EEL spectra is 0.75eV under normal operation conditions. Our EELS experiments were performed with the convergence angle of ~0.7mrad and the spectrometer collection angle of ~3.0mrad.

## 3. Results and discussion

### 3.1. Band structure

Fig.1 shows the calculated LSDA band structure of $Sr_{0.35}CoO_2$ near Fermi level ($E_F$): (a) spin up electrons and (b) spin down electrons. The crystal-field split of the Co 3d bands ($e_g$-$t_{2g}$) is estimated to be around 2eV in present system, which is smaller than that (~2.5eV) in $Na_xCoO_2$ materials [13]. The unoccupied $e_g$ states with a narrow bandwidth of around 0.9eV appear at 1.2eV above $E_F$; the $t_{2g}$ states with the bandwidth of ~1.8eV are intersected by $E_F$. The occupied O 2p bands extend from -7eV to -1.5eV. The $t_{2g}$ manifold splits into $a_{1g}$ and $e'_g$ bands in the rhombohedral crystal field as discussed in $Na_xCoO_2$ (0<x<0.7) materials [13]. The primarily $a_{1g}$ character is emphasized with dark-circles proportional to its amount. The $a_{1g}$ band has almost the same bandwidth as the $t_{2g}$ manifold. In comparison with the band structure of $Na_xCoO_2$ materials, the $a_{1g}$ bands in $Sr_{0.35}CoO_2$ are noticeably depressed at Γ(k=0) point, and partially go down $E_F$. They also show apparent dispersions along the Γ-A line in sharp contrast to the flat bands with little c-axis dispersion in $Na_xCoO_2$ system [13]. Fig. 2 shows the LSDA Fermi surface (FS) for $Sr_{0.35}CoO_2$, which was calculated on a 45×45×9 mesh containing 960 k-points in the irreducible BZ and viewed by XCrySDen package [18]. The Fermi level crosses both the $a_{1g}$ and one of the $e'_g$ bands; the $t_{2g}$ manifold produces a small and a large Fermi surfaces around the Γ point; the $e'_g$ band produces six pockets near K points in the Brillouin zone (BZ) in the direction of K-Γ. The remarkable feature of FS topology, notably different from $Na_xCoO_2$, is the six small pockets showing up as holey ellipsoids containing $e'_g$ holes instead of six small cylinders in $Na_{0.5}CoO_2$ [13]. This alternation results from the apparent increase of c-dispersions of the $t_{2g}$ bands in $Sr_{0.35}CoO_2$. Integration of unoccupied states in the density of states (DOS) give rise to 0.28 holes/Co to account for the $Sr_{0.35}CoO_2$ phase in which two 5s electrons arise from per Sr atom.

Previously, we have performed a series of experimental measurements on $Sr_xCoO_2$ materials. It is found that the transport property of the $Sr_{0.35}CoO_2$ polycrystalline samples depends sensitively on annealing conditions; their resistivities in general are metallic [8, 11] in agreement with the half-metallic band structure as illustrated in fig.1. Experimental measurements on the magnetic properties of $Sr_{0.35}CoO_2$ indicate that magnetic susceptibility ($\chi$) increases with lowering temperature and obeys the Curie-Weiss law. Above 50 K, $\chi$ can be well fitted by the formula $\chi(T)=C/(T-\theta)$, with $\theta$ = -133.1K, The negative value of $\theta$ indicates that the spin correlations are antiferromagnetic (AFM) . The fitting parameters allow us to give rise to the effective moments as 1.36$\mu_B$/Co in $Sr_{0.35}CoO_2$, which is close to that of $Na_xCoO_2$ with x ~0.7 antiferromagnetic spin correlations above 50K [11]. In order to understand the observed antiferromagnetic correlation, we have performed a further calculation based on a simple AFM model by flipping one of the Co spin (two Co atoms/cell) to form an interlayer AFM structure with a space group of P-3m1. Our LSDA calculation indicated that the ground state of this model tends to converge at a FM state rather than the expected AFM state. Hence, a larger unit cell may be needed in further study of magnetic properties.

It is generally agreed that the Coulomb correlation should be concerned for the understanding of physical properties in $Na_xCoO_2$ system, the correlation effects and strength could vary with the doping level [19]. The $Sr_{0.35}CoO_2$ material also has a narrow $t_{2g}$ band with bandwidth W=1.8eV in comparison with the estimated U for Co (5-8eV) [13]. We therefore employed the LSDA+U method to study the correlated effect on electronic band structure in present system with U ranging from 0 to 8eV. Fig. 3 shows the density of states (DOS) for four typical U values, the partial DOS of Co 3d and O 2p states are respectively plotted by dashed line and dotted line. It is found that the electronic structure is sensitive to the variation of U. The spin-down unoccupied $a_{1g}$ states split progressively from the valence band along with the U increase from 0 to 3eV. As U is larger than 3eV, the spin-up $e_g$ bands shift noticeably toward the lower energy direction, and on the other hand, the spin down $t_{2g}$ bands shift in opposite direction toward the higher energy region. As a result, additional unoccupied $t_{2g}$ states appear above $E_F$.

3.2. Comparison between theoretical and experimental EELS

Electron energy-loss near-edge structure provides the unoccupied partial DOS around the excited atom, so we performed a series of experimental EELS measurements on $Sr_{0.35}CoO_2$ and compared with the theoretical results. Fig. 4 (a) and (b) show respectively the oxygen K-edge and Co $L_{23}$-edge core loss EELS after background subtraction for $Sr_{0.35}CoO_2$, both experimental data and numerous theoretical results are displayed for comparison. The resolutions of all calculated spectra are lowered so that the major peaks have the same full-width at half-maximum (FWHM) as the experimental ones. Analysis of spectral features for different Coulomb interactions (U) suggests that critical alternations for either the O K-edge or the Co $L_{23}$-edge appear at around U=3eV; peak splitting can clearly visible in all spectra for U≥4eV.

The oxygen K-edge spectra gives the oxygen p-projected DOS of ground state and the core hole potential does not have large effect on the spectral shape. The five main peaks as shown in fig.4 (a) can be well-interpreted base on our theoretical analysis. Peak *a* appears at 1.6eV above Fermi level, representing the transitions from O 1s towards the unoccupied O 2p mixed with Co $e_g$. Peak *b*, at ~ 6.5eV above peak *a*, arises from a combination of transitions from O 1s to Sr 4p, Co 4p and O 2p. Peak *c*, at ~12eV above peak *a*, is much broader in comparison with peak *a* or *b*, reflecting the transitions from O 1s to the unoccupied states mixed by Co 4s4p, Sr 4s4p and O 2p. Peak *d* and *e* in general are very broad as observed in other transition metal oxides, which correspond to the mixture of Co 4p3d, Sr 4p3d and O 2p. Previously, these broad peaks have been also analyzed in connection with the single-scattering events from different oxygen coordination shells as pointed out by H. Kurata [20]. It is noted that the calculated spectra for U≤3eV can perfectly reproduce all peaks appearing in the experimental spectra, their positions, as well as intensities, are fundamentally in good agreement with the experimental ones. The most striking feature in the theoretical O K-edge spectra is the appearance of the remarkable split on the first peak *a* for U≥4eV. This phenomenon results essentially from evolution of band structures along with the increase of U. The shifts of the spin up and spin down bands occur in opposite directions and yields additional unoccupied spin down $t_{2g}$ states above $E_F$ for U≥4eV. Hence, previous to the peak *a*, another new peak becomes visible reflecting the transitions to the new unoccupied states. These two peaks with around 2.2eV apart should be visible under our experimental resolution. However, no evidence for peak splitting has been found in our careful observations, we therefore conclude that the on-site interaction (U) in $Sr_{0.35}CoO_2$ is less than 3eV. This conclusion is all supported by a further analysis of the Co $L_{23}$-edge as demonstrated in fig. 4(b), in which peak *a* and *b* reflect respectively the transitions from Co $2p^{3/2}$ and Co $2p^{1/2}$ to unoccupied Co 3d states. EELS calculations for U≥4eV reveal again the peak splitting for both $L_3$ and $L_2$ transitions. The experimental spectrum is qualitatively in consistence with theoretical data for U≤3eV. Hence, $Sr_{0.35}CoO_2$ is a moderately correlated system, the strength of the on-site coulomb interaction is notably smaller than that estimated for $Na_xCoO_2$ (5-8eV) [13, 16]. It is also possible that the correlation effect/strength could vary with the doping level due to the change in electron screening as discussed in $Na_xCoO_2$ [19].

## 4. Conclusions

In conclusion, the band structure and Fermi surface topology of $Sr_{0.35}CoO_2$ have been studied using LSDA. The Co 3d bands show up evident c-dispersions arising from notable interplanar interaction in $Sr_{0.35}CoO_2$. These results suggest that $Sr_{0.35}CoO_2$ should be much more three-dimensional than the $Na_xCoO_2$ materials. The LSDA+U calculation reveals that the electronic structure, as well as the split between the spin-up and spin-down bands, changes remarkably along with the U increase, especially for U$\geq$4eV. Systematical analysis of the theoretical and experimental EELS for O K-edge and Co L$_{23}$-edge suggests that the electronic correlation U in this system is less than 3eV.

**Acknowledgments**

We would like to thank Prof. B. G. Liu and Dr. L. J. Shi for their valuable discussions. The work reported here is supported by 'National Natural Foundation' and 'Outstanding Youth Fund' of China.

References

1. I. Terasaki, Y. Sasago and K. Uchinokura, Phys. Rev. B **56** (1997) 12685.

2. T. Kawata, Y. Iguchi, T. Itoh, K. Takahata and I. Terasaki, Phys. Rev. B **60** (1999) 10584.

3. T. Motohashi, E. Naujalis, R. Ueda, K. Isawa, M. Karppinen and H. Yamauchi, Appl. Phys. Lett. **79** (2001) 1480.

4. K. Fujita, T. Mochida and K. Nakamura, Jpn. J. Appl. Phys. **40** (2001) 4644.

5. K. Takada, H. Sakurai, E. Takayama-Muromachi, F. Izumi, R.A. Dilanian and T. Sasaki, Nature **422** (2003) 53.

6. M.L. Foo, Y.Y. Wang, S. Watauchi, H.W. Zandbergen, T. He, R.J. Cava and N.P. Ong, Phys. Rev. Lett. **92** (2004) 247001.

7. F.C. Chou, J.H. Cho and Y.S. Lee, Phys. Rev. B **70** (2004) 144526.

8. R. Ishikawa, Y. Ono, Y. Miyazaki and T. Kajitani, Jpn. J. Appl. Phys. **41** (2002) L337.

9. B.L. Cushing and J.B. Wiley, J. Solid. State. Chem. **141** (1998) 385.

10. H.X. Yang, C.J. Nie, Y.G. Shi, H.C. Yu, S. Ding, Y.L. Liu, D. Wu, N.L. Wang and J.Q. Li, Solid. State. Comm. **134** (2005) 403.

11. H.X. Yang, Y.G. Shi, Y.Q. Guo, X. Liu, R.J. Xiao, J.L. Luo and J.Q. Li (submitted).

12. A. Van der Ven, M.K. Aydinol, G. Ceder, G. Kresse and J. Hafner, Phys. Rev. B **58** (1998) 2975.

13. D.J. Singh, Phys. Rev. B **61** (2000) 13397；K.-W. Lee, J. Kuneš and W. E. Pickett, Phys. Rev. B **70** (2004) 045104.

14. P. Blaha, K. Schwarz, G.K.H. Madsen, D. Kvasnicka and J. Luitz, WIEN2k, An

Augmented Plane Wave + Local Orbitals Program for Calculating Crystal Properties (revised edition June 2002) (Karlheinz Schwarz, Techn. Universität Wien, Austria) ISBN 3-9501031-1-2.

15. P. Novák, F. Boucher, P. Gressier, P. Blaha and K. Schwarz, Phys. Rev. B **63** (2001) 235114.

16. H. Okabe, M. Matoba, T. Kyomen and M. Itoh, J. Appl. Phys. **95** (2004) 6831.

17. M. Nelhiebel, P.-H. Louf, P. Schattschneider, P. Blaha, K. Schwarz and B. Jouffrey, Phys. Rev. B **59** (1999) 12807.

18. A. Kokalj, XCrySDen—a new program for displaying crystalline structures and electron densities, J. Mol. Graphics Modelling **17** (1999)176. Code available from http://www.xcrysden.org/.

19. K.-W. Lee, J. Kuneš, P. Novak and W. E. Pickett, Phys. Rev. Lett. **94** (2005) 026403.

20. H. Kurata, E. Lefèvre, C. Colliex and R. Brydson, Phys. Rev. B **47** (1993) 13763.

Figure Captions

**Fig. 1** The LSDA band structure for $Sr_{0.35}CoO_2$ near the Fermi level: (a) spin up bands; (b) spin down bands. The horizontal line indicates the Fermi level. The $a_{1g}$ character is emphasized with circles proportional to its amount.

**Fig. 2** The LSDA Fermi surface of $Sr_{0.35}CoO_2$ in the (a) $k_z=0$ and (b) $k_z=0.5$ planes. Shaded areas are electron state; the six small pockets contain holes that are primarily $e'_g$-like.

**Fig. 3** DOS of $Sr_{0.35}CoO_2$ for (a) LSDA (U=0), (b) U=3eV, (c) U=4eV and (d) U=8eV. Solid curves denote the total DOS, dashed lines denote DOS of Co 3d orbitals, and dotted lines denote DOS of O 2p orbitals.

**Fig. 4** Experimental and calculated EEL spectra of (a) O K-edge and (b) Co $L_{23}$-edge. The calculated spectra are broadened to the same resolution as the experimental ones following Gaussian function.

Fig. 1

(a)

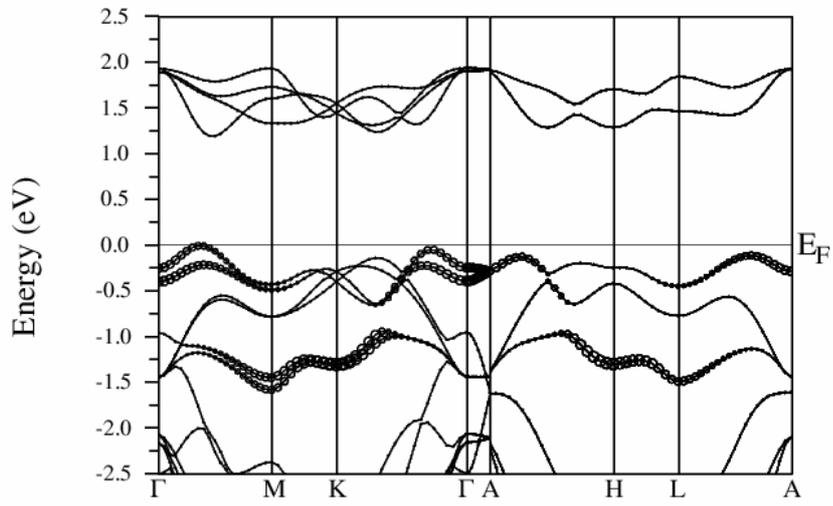

(b)

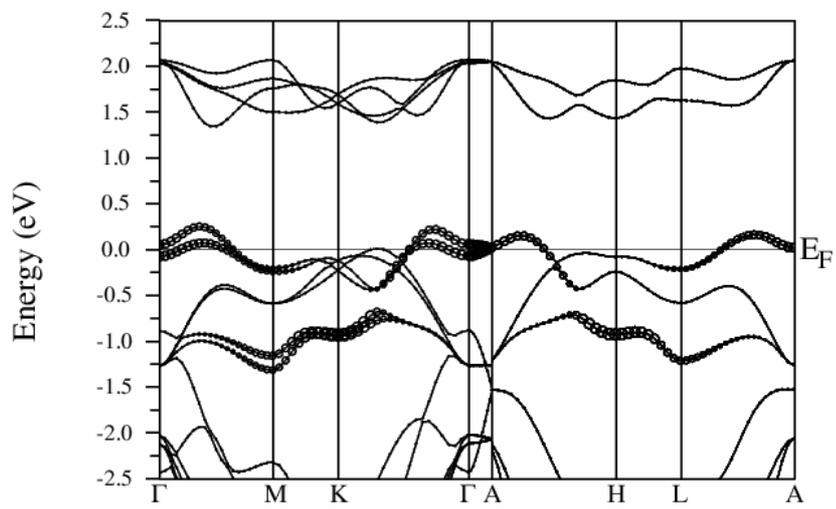

Fig. 2

(a)

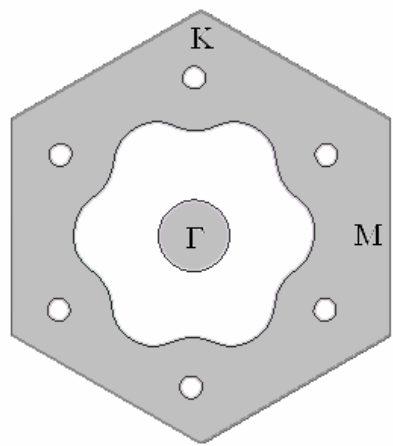

(b)

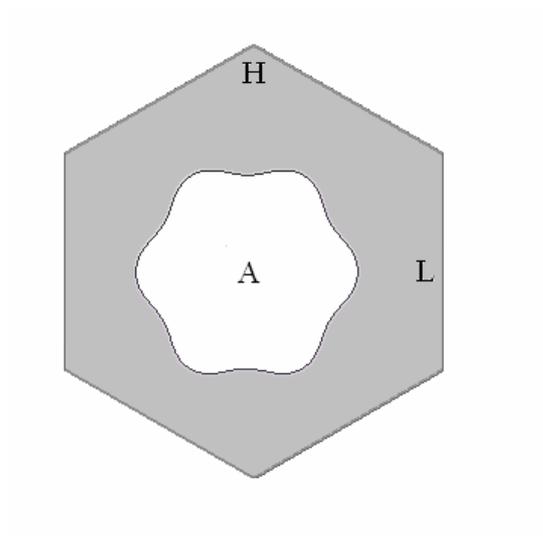

Fig. 3

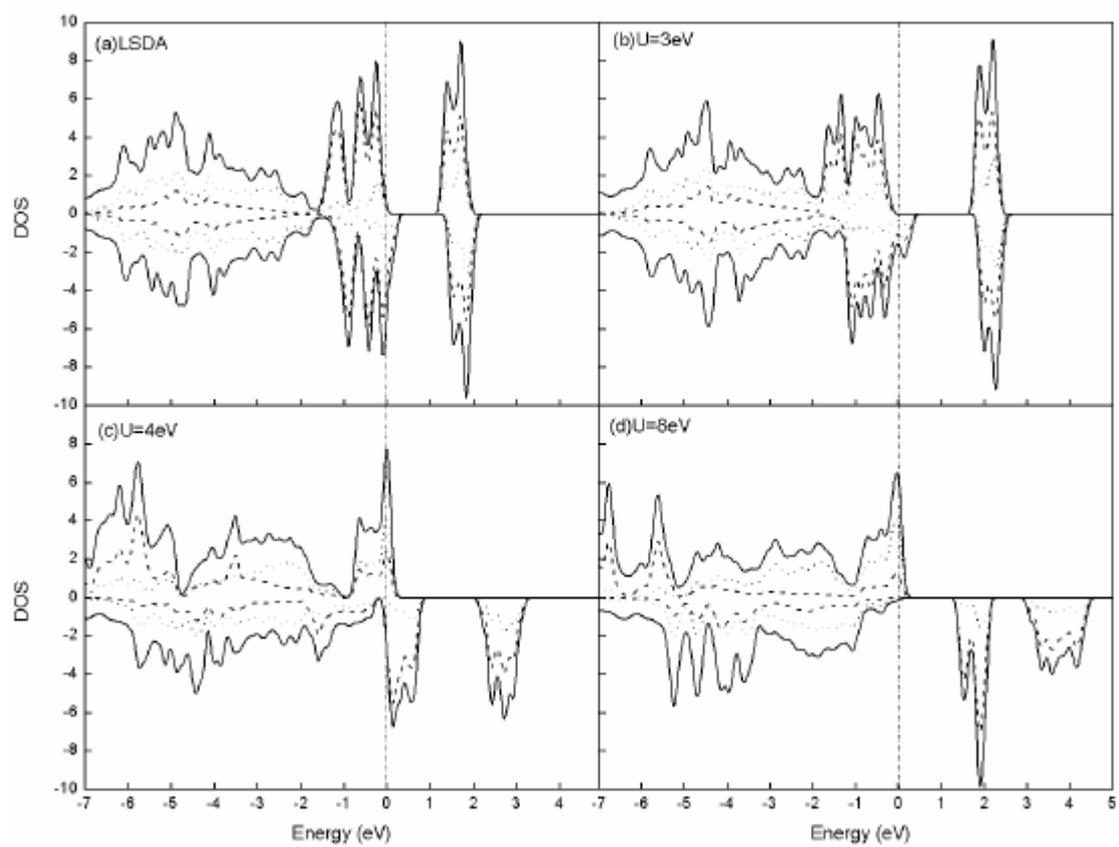

Fig. 4

(a)

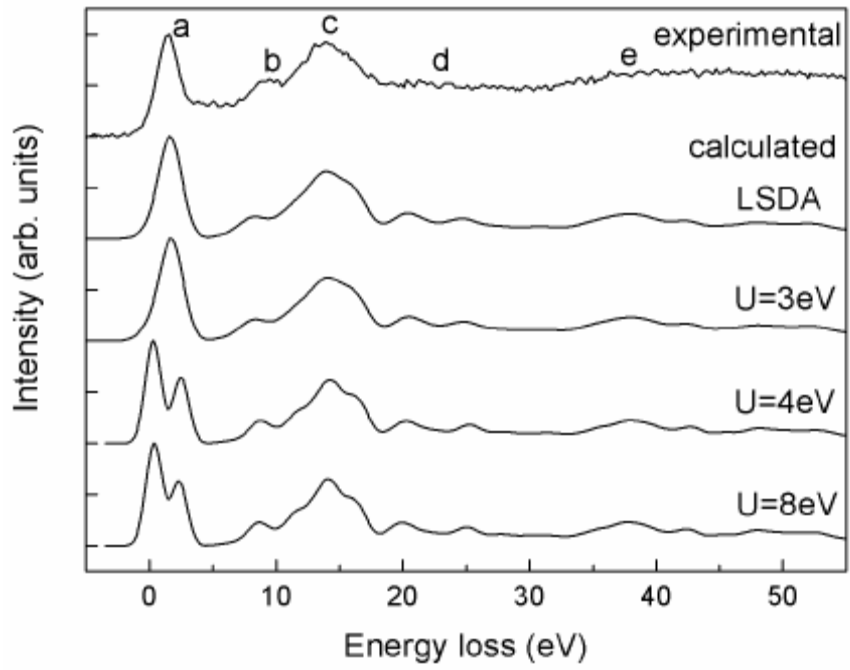

(b)

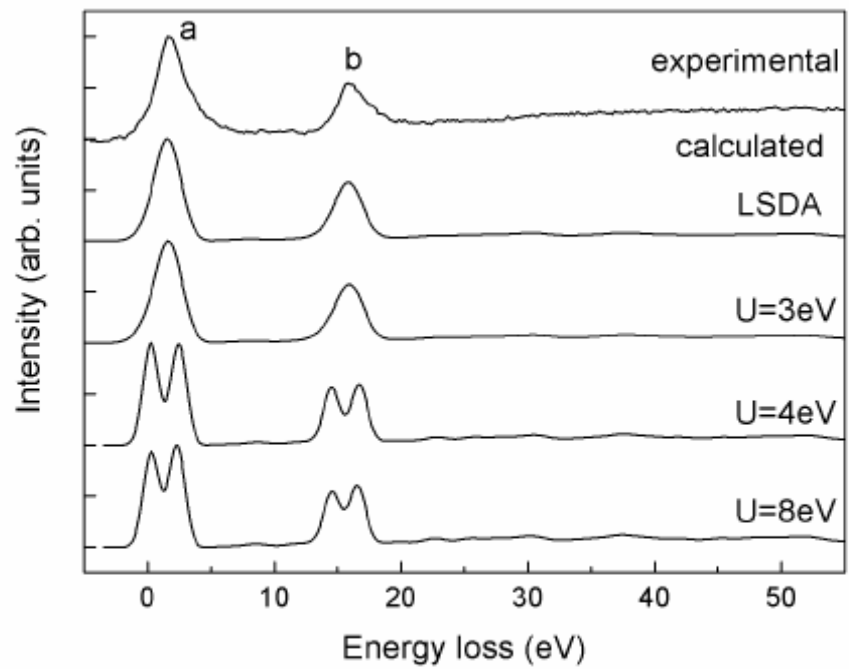